\documentclass[12pt,preprint]{aastex}

\shorttitle{h and Chi Persei}
\shortauthors{Slesnick et al.}

\def\hchi{h/$\chi$ Per~} 
 
\def\hPer {h Per~} 
\def\ChiPer {$\chi$ Per~} 
\def\mbol{M$_{bol}$~}
\def\logt{log T$_{\rm eff}$~}

\begin{document}

\title{The Star Formation History and Mass Function of the Double Cluster h and Chi Persei}

\author{Catherine L. Slesnick\altaffilmark{1}, Lynne
A. Hillenbrand\altaffilmark{1}}

\affil{Dept.\ of Astronomy, MS105-24, California Institute of
Technology,Pasadena, CA 91125}

\email{cls@astro.caltech.edu, lah@astro.caltech.edu}
\and
\author{Philip Massey\altaffilmark{1}}
\affil{Lowell Observatory, 1400 W. Mars Hill Road, Flagstaff, AZ 86001}
\email{Phil.Massey@lowell.edu}

\altaffiltext{1}{Visiting astronomer, Kitt Peak National Observatory,
a division of the National Optical Astronomy Observatory, which is
operated by the Association of Universities for Research in Astronomy,
Inc., under cooperative agreement with the National Science Foundation.}

\begin{abstract}

The h and $\chi$ Per ``double cluster" is examined using wide-field
($0.98^\circ \times 0.98^\circ$) CCD {\it UBV} imaging supplemented
by optical spectra of several hundred of the brightest stars.  Restricting our
analysis to near the cluster nuclei, we find identical reddenings
($E(B-V)=0.56\pm0.01$), distance moduli ($11.85\pm0.05$), and ages 
($12.8\pm1.0$~Myr) for the two clusters.
In addition, we find an IMF slope for each of the cluster
nuclei that is quite normal for high-mass stars, $\Gamma=-1.3\pm0.2$,
indistinguishable from a Salpeter value.  We derive masses of
3700 $\cal M_\odot$ (h) and 2800 $\cal M_\odot$ ($\chi$)
integrating the PDMF from 1 to 120 $\cal M_\odot$. 
There is evidence of mild mass segregation within the cluster cores.
Our data are consistent with the stars having formed at a single
epoch; claims to the contrary are very likely due to the inclusion of the
substantial population of early-type stars
located at similar distances in the Perseus
spiral arm, in addition to contamination by G and K giants at various
distances.  We discuss the uniqueness of the double cluster, citing
other examples of such structures in the literature, but concluding
that the nearly identical nature of the two cluster cores is unusual.  We fail
to settle the long-standing controversy regarding whether or not the double cluster is the
core of the Per OB1 association, and argue that this may be unanswerable
with current techniques.  We also emphasize the need for further work on
the pre-main sequence population of this nearby and highly interesting
region.

\end{abstract}


\section{Introduction}

The ``double cluster'' h and $\chi$ Persei (hereafter h/$\chi$ Per; also
known as NGC 869 and NGC 884, respectively) is among the brightest,
densest, and closest of the open clusters containing moderately massive
stars.  The double cluster has been studied extensively over the last century (e.g. Oosterhoff 1937;
Bidelman 1943; Wildey 1964; Schild 1965, 1967; Crawford, Glaspey, \&
Perry  1970; Vogt 1971; Tapia et al.\ 1984; Waelkens et al.\ 1990) with
resulting mean reddenings of $E(B-V)$ = 0.5-0.6, and distance moduli in
the range 11.4-12.0 mag (1.9-2.5~kpc).  The clusters contain several tens
of Be stars (e.g. Slettebak 1968; Bidelman 1947a; see also Keller et al.\ 2001).  Wildey's (1964) HR
diagrams suggested several distinct episodes of star formation (7~Myr,
17~Myr, and 60~Myr), which would imply a spread of $>$50 Myr in the
formation times of OB stars in a single (double) cluster!  This age spread
is larger than that claimed for any other well-studied open cluster,
and is one of the primary motivations of the present investigation.

Most previous work on \hchi has used photographic or single-channel
photoelectric photometry with little emphasis on spectroscopy.  Several
very recent papers have used CCDs but consisted of photometric 
analysis only ($UBVI$/H$\alpha$, Keller et al.\ 2001; $ubvy/\beta$, Marco
\& Bernabeu 2001).  Distance moduli in the range 11.6-11.8 mag and ages
of 10-20~Myr have been found, with Marco \& Bernabeu (2001) arguing
(like Wildey 1964) for three distinct episodes of star formation, while
Keller et al.\ (2001) find instead a single age.  
There is significant disagreement between
various authors as to whether the reddenings,
distances, and ages of the two clusters are identical or substantially
(30-50\%) different.  It is especially important to understand in detail
the star formation history of \hchi as these clusters are widely used
from professional review papers to basic introductory astronomy textbooks
to illustrate upper main sequence stellar evolution.

Our modern study of \hchi consists of wide-field CCD {\it UBV}
photometry for 4528 stars
and blue optical spectroscopy for 196 of the stars presumed to be
the most massive (i.e., the brightest blue and red stars).
Our main goals are to re-determine the distance, age, and age spread
in the double clusters,  and to explore for the first time the mass 
function and the evidence for mass segregation.  
In section 2 we describe our data acquisition, 
reduction, and preliminary photometric and spectroscopic analysis.  
In section 3 we present color
magnitude diagrams, an assessment of field star contamination, a derivation
of reddening and distance, and a Hertzsprung-Russell (HR) diagram 
along with discussion of stellar age and mass distributions. Section 4 
contains our discussion of the uniqueness of this double cluster, and the
relationship between it and the surrounding region.  In Section 5 we
summarize our results.

\section{Observations and Data Reduction}

\subsection{Photometry} 

{\it UBV} photometry 
was obtained from observations with the 0.9-m telescope
at Kitt Peak National Observatory using the Mosaic CCD camera 
(0.43 arcsec/pixel) on 1999 Feb 3.  Conditions were photometric
with $\sim$1.3 arcsec seeing.
The Mosaic camera consists of
8 individual SITe $2048 \times 4096$ CCD chips arranged 
in 2 rows of 4 to produce a 
final image equivalent to 8192 $\times$ 8192 pixel$^2$ (0.98 $\times$ 0.98 deg$^2$) 
but with modest (35 to 50 pixel) gaps.
Our imaging data set contains short (0.5 sec in $V$ and $B$,
and 2 sec in $U$), medium (2 sec in $V$ and $B$, and 10 sec in $U$), 
and long (100 sec in $V$ and $B$, and 300 sec in $U$) integrations, 
each consisting of 5 dithered exposures which were combined 
to fill in gaps between the 8 chips.  
The exceptions to this pattern are the short
exposures which were not dithered, and the medium $B$ exposure which only had 
4 dithers instead of 5.  Many Landolt (1992) standards were observed for
the purposes of transformation to the {\it UBV} system.

For the basic reductions we followed the precepts of Valdes (1998),
using the IRAF ``mscred" package.  With bright twilight flats we were
able to flatten the data to $<1\%$ in terms of large-scale gradients.
As the plate scale changes significantly over the field of view, care must be
taken to geometrically correct the data to a uniform sampling for the
premise of aperture photometry to work; this transformation was made
using a sinc interpolation in order to come close to preserving 
the Poisson noise characteristics.

A common but hitherto untested  practice when working with Mosaic data
is to combine each set of ditherings into a single ``stacked" image
for photometry.  However, each chip has its own spectral response and
hence, color term.  Since any given star may be the average of multiple
dithered exposures and may appear on up to 4 of the CCDs, we were driven
to wonder to what degree of accuracy one could do photometry using
the final combined images.  We thus performed both aperture photometry
and point-spread-function (PSF) fitting photometry separately on both
the stacked images (9 total = 3 colors $\times$ 3 integration times)
and the individual CCD frames (247 total = 8 chips $\times$ 1,4, or 5
ditherings $\times$ 3 colors $\times$ 3 integration times).  Color terms
were determined for each of the 8 chips individually.  We retained the
median value in each filter for the stacked images.  On the
whole, color-term variations were most significant at $U$.  By adopting
a single color-term for the stacked images we expect to make systematic
errors in $V$ by an average of 0.012 magnitudes over a color range in
$(B-V)$ of 1.  The maximal chip to chip difference is 0.032 mag in $V$.
The $B$ band gave similar residuals with a full range of 0.038 per
one magnitude range $(B-V)$.  In $U$ band the chip-to-chip offsets were
considerably larger, yielding typical variations of 0.038 mag over a
range of 1 mag in $(U-B)$ and a maximal difference of 0.11 mag.

We found that PSF-photometry of the stacked images produces errors
of $> 5\%$, independent of magnitude, suggesting that these errors
are not dominated by photon noise.  The scatter for the individual
frame PSF photometry was also magnitude-independent and gave errors of
$\sim3\%$ when compared to aperture photometry of single isolated stars.
We could see by visual inspection that there were significant variations
in the PSF even across a single chip, despite the relatively slow
(f/7.5) beam.\footnote{Subsequent to these data being obtained, the
corrector in the 0.9-m was realigned, leading to improved behavior of
the PSF.} Accordingly we choose to rely upon aperture photometry alone,
sacrificing the potential advantage of PSF-fitting for any crowded stars.
Fortunately \hchi is relatively sparse.
Over the entire imagin area, the mean stellar density 
to $V=15$~mag is 0.7 stars arcmin$^{-2}$; in the center of the clusters,
it is still a modest
2.2 and 1.7 stars arcmin$^{-2}$ for h and
$\chi$, respectively.

Our standard and program stars were each measured with the same
large aperture (10 pixel radius = 8.6 arcsec diameter). The standard star
data were then used to produce transformations between the instrumental
and standard system, The extinction values we found were typical of Kitt
Peak, and our fits had small ($<0.02$~mag) residuals.

Because our frames went much deeper than any program stars of interest
(thanks to the long-exposure frames) we had the luxury of retaining only the
very best data for the subsequent analysis.  After merging the data for
the three sets of exposures times, we kept only those stars for which
the instrumental errors (due to phone-statistics and read-noise) were
less than 0.01 mag in {\it U}, {\it B}, and {\it V}. This eliminated roughly
96\% of the stars we had measured, and is equivalent to simply imposing a 
magnitude cut-off on the data.  Thus while our catalog does not go as deep
as other recent efforts (cf. Keller et al 2001) our photometric errors are
quite small and we purposefully chose to truncate our catalog once
field contamination became extreme.

Table 1 contains our catalog of \hchi stars, ordered by decreasing brightness.
We have merged the photometry for the three sets of exposures, weighting
inversely by the square of the photometric uncertainty.
Our final
source list contains
three-filter photometry for 
4528 stars down to $V$ $\sim$16 mag.  We can estimate our completeness
from the histograms of the number of stars per 0.25~mag bin shown in 
Fig. 1 and find completeness of our catalog
to $U\sim16.2$, $B\sim 16.0$, $V\sim 15.0$.

How well does our photometry agree with earlier studies?  
In Fig. 2 we compare our work with a subset of Wildey's (1964)
photoelectric and photographic work, concentrating primarily on the brighter 
stars and those for which we have spectral types.  
We see that his photometry and ours agree
extremely well given the differences in equipment.  We find 
differences (in the sense of our data minus Wildey's) with his photographic data 
of ($\Delta$$V$)$_{avg}$ = 0.097$\pm$0.027, [$\Delta(B-V)$]$_{avg}$
 = -0.044$\pm$0.009, and [$\Delta(U-B)$]$_{avg}$
 = -0.069$\pm$0.036 computed from a comparison of 300 stars. Comparing our 
photometry to 24 of Wildey's photoelectrically observed stars we find even  
smaller average offsets 
of ($\Delta$$V$)$_{avg}$ = 0.020$\pm$0.022, [$\Delta(B-V)$]$_{avg}$
 = -0.014$\pm$0.026, and [$\Delta(U-B)$]$_{avg}$
 = 0.017$\pm$0.012.
The scatter is larger and asymmetrical in the $V$ comparison (see Fig. 2),
in the sense one would expect if Wildey's work occasionally had
faint stars in the sky determination.  
In a similar comparison with Wildey's photographic study, Keller et al. (2001) 
quote average differences of 
($\Delta$$V$)$_{avg}$ = 0.16, [$\Delta(B-V)$]$_{avg}$ = -0.03, 
and [$\Delta(U-B)$]$_{avg}$ = 0.00, attributing
the offset with respect to Wildey's $V$-band photometry to crowding effects.  
We have matched our data to that of Keller et al.\ (2001) for stars 
which are not known from the litterature to be variable 
and for which we have identifed 
Oosterhoff (1937) numbers (again, our cross-identification is not complete),
 and find average offsets of 
($\Delta$$V$)$_{avg}$ = -0.019$\pm$0.008, [$\Delta(B-V)$]$_{avg}$
 = 0.001$\pm$0.026, and [$\Delta(U-B)$]$_{avg}$
 = -0.048$\pm$0.014 computed from 55, 31 and 49 stars, respectively.    
The generally good agreement between all three studies
is a testament to Wildey's painstaking accuracy
in centering stars with a photoelectric photometer, and also supports
the validity of our reduction procedure and transformation to the standard
photometric system.

\subsection{Spectroscopy} 

Several hundred spectral types complement our photometric database.
For hot stars, spectral data are needed to obtain accurate effective
temperatures and consequently accurate extinction estimates and bolometric
corrections (Massey 1998a, 1998b), all necessary for locating a star in the HR
diagram.  We selected stars for spectroscopy based on their brightness
and colors.  Since we did not yet have our own CCD photometry at the time
the spectroscopic program was begun we worked largely from the Wildey
(1964) photometry; this introduces a bias towards blue stars closer to
the cluster nuclei.  Later spectroscopic runs incorporated a wider range
of magnitude and color selection, probing down as far as mid-A spectral
types in an unrealized attempt
to identify possible pre-main sequence stars amidst
substantial field star contamination.  In the lower-left panel of
Fig. 4 we show the spatial distribution of the spectroscopic
sample compared to the entire photometric sample; in the upper-left panel
we show the loci of the spectroscopic sample in the color-magnitude plane.
Of the brightest 50 stars, we have spectral types for 49, regardless of
location in our field.  Fainter than that the spectroscopic campaign was
concentrated in the cluster cores.

Spectroscopic data were taken at several NOAO telescopes.  We employed
the WIYN 3.5-m telescope and the Hydra multi-fiber positioner to feed a
bench-mounted spectrograph (1993 Dec and 1994 Oct), the KPNO 4-m telescope
and the RC spectrograph in multi-slit mode (1994 Nov and 1999 Aug),
the KPNO 2.1-m telescope and GoldCam with a single slit (1994 Sept),
and the Coude Feed (1999 July and Nov).  For most of the spectra the
spectral range is $\sim\lambda\lambda3900-4700$ \AA~ at a resolution
of $\sim$1.5 \AA.  Higher resolution was obtained with the Coude Feed
data, which were taken in multiple wavelength settings.  One-dimensional
spectra were extracted from the two-dimensional images using the slit
and multi-fiber reduction packages within IRAF.  Signal-to-noise ranged
from $\sim$20-150 with nearly all spectra classifiable.  Fig. 3
shows three spectra taken at the KPNO 2.1-m telescope with GoldCam
which illustrate the effects of temperature on B-type supergiants.

We present in Table 1 new spectral types for 196 stars, many of
which result from several different observations of the same object.
In classifying the spectra we followed the guidelines of Walborn \&
Fitzpatrick (1990), Jaschek \& Jaschek (1987), and Jacoby, Hunter, \&
Christian (1984).  All stars were classified by a minimum of two of the
authors, both independently and collaboratively.  Spectral types assigned
by us were also compared to those in the literature where available,
and we cite these older spectral types as well.  As emphasized in the
introduction, spectroscopic efforts have lagged behind photometric studies
of this region.  Johnson \& Morgan (1955), 
Schild (1965, 1967), and Slettebak (1968) have made the most systematic
efforts in this regard, and in general our spectral types agree very
well with theirs.

The most luminous stars we identify in the vicinity of \hchi are M, A,
and B supergiants.  There is a lone O-type star, HD~14434 (O6.5V). As
we discuss below, this star is likely not a member of the double cluster,
but appears to be a younger field star interloper at approximately the
same distance.  The remainder of the spectra are slightly evolved B-type
giants and B- and A-type dwarfs.  We identify ten Be stars, two of which
were previously unknown.  Six out of seven stars which were classified photometrically
as Be stars by Keller et al.\ (2001) using $(V-H\alpha)$ colors 
and which we have our own spectra for, do in fact 
prove to be emission line objects.  Since none of our spectra extend as far 
redwards as H$\alpha$, emmision seen by us is usually in H$\beta$ 
which tends to be weaker than H$\alpha$ emission by $\approx$1/3.

\section{Analysis}

Our analysis includes discussion of color-magnitude diagrams, assessment
of field star contamination, derivation of cluster reddening and distance,
and construction of HR diagrams.  We then discuss stellar ages and masses
as inferred from the HR diagrams, and the distribution of ages and
masses within the clusters.

We expect that stars near the cluster cores predominantly will be members,
while stars further afield will be a mixture of both members and non-members.
We constructed a contour plot of the spatial distribution of stars within our field,
and found that the stellar densities were enhanced by 2$\sigma$ at identical radii
of 7 arcminutes from each of the cluster cores; we will use this radial criterion
when describing stars near the nuclei.
We also determined accurate centers for the two cores
($\alpha_{2000}$=2:19:22.2,
$\delta_{2000}$=+57:09:00 for \hPer, and 
$\alpha_{2000}$=2:22:12.0,
$\delta_{2000}$=+57:07:12 for \ChiPer) 
by examining mass and number density contours.

Tables~2a \&~2b contain derived quantites for stars near to ($\leq 7$ arcmin) and further away 
from the cluster centers, respectively.  
We have limited these tables to those stars which we included in 
determing the PDMF (see section 3.4.4).

\subsection{Color-Magnitude Diagrams and the Influence of Field Stars}

In Figure 4 we show the color-magnitude
diagram (CMD) for (1) 
all of the stars over the full 0.98 $\times$ 0.98 deg$^2$ covered in
our CCD images (left panels), and (2) only the stars within 7 arcmin
of each of the two cluster nuclei. 
The influence of field stars
can be seen in the upper left panel of this figure, notably between $0.5
< (B-V) < 1.0$ and $10 < V < 16$.  For the h and $\chi$ cluster nuclei
in the upper right panel, field star contamination is less severe, but
still present.  The close match between the two CMDs suggests that there
are no substantial differences in reddening, distance or age between
the two clusters, a conclusion we explore in greater depth below.

To further assess the field star contamination we use the density
of stars in the color-magnitude diagrams (i.e., a ``Hess" diagram),
as shown in Figure 5 for both $V$ {\it vs.} $(B-V)$ and $V$
{\it vs.} $(U-B)$.  We define the ``cluster region" as above, while
the ``field stars'' region is arbitrarily taken as
the northern 0.25 deg and southern 0.15 deg of our CCD imaging area.
Hess diagrams produced for this total area of 0.4 deg$^2$ were scaled up
to the full area of the CCD survey (shown as red contours in the upper
panels of Figure 5) and subtracted from the Hess diagram
constructed over the full imaging area.  The result of this subtraction
is shown in the lower panels, with the resulting $V$ {\it vs.} $(B-V)$ and $V$
{\it vs.} $(U-B)$ diagrams displaying a much tighter color-magnitude sequence
than the full area.

\subsection{Reddening}
	
For each star with a spectral type we compute the color excess using
the spectral type-intrinsic color relations of FitzGerald (1970).
The average value of color excess for 123 stars with well determined spectral
types is $E(B-V)$ = 0.55, with a 1$\sigma$ variation of 0.1.   For the
56 stars near the core of h Per we find an average $E(B-V)=0.57$, with
a 1$\sigma$ variation of 0.08.  Similarly for 40 stars near 
the core of $\chi$~Per we
find an average of 0.53 (1$\sigma$=0.08).  The median values are 0.56~mag,
0.57~mag, and 0.55~mag for the three samples, respectively.  We conclude 
that the reddening is indistinguishable for the two clusters and, further,
infer that the reddening is entirely due to line of sight extinction to the
Perseus spiral arm with intracluster reddening essentially zero.
This is consistent with a 3$\sigma$ upper limit on $^{13}$CO emission
(Hillenbrand \& Carpenter; unpublished FCRAO data) corresponding to 
essentially no gas ($N(^{13}CO) < 2.1 \times 10^{15}$ cm$^{-2}$) or dust 
(A$_V <$ 1.4 mag) within the clusters.

In dereddening the photometry, we assume the standard 3.1 ratio
of total-to-selective absorption.
For other stars without spectral
types, color excesses were determined using the ``$Q$ method", where
applicable: following Massey (1998a), for stars with $B-V<0.5$,  we compute
$(B-V)_o=-0.0186+0.3218 \times Q$, where $Q=(U-B)-0.72 \times (B-V)$ is
a reddening-independent index.  The relation between $(B-V)_o$ and $Q$
was derived by fitting the intrinsic color relationships of FitzGerald
(1970) for main-sequence stars, but the relationship can be used for
supergiants and giants.  For stars without spectral types, we adopt 
the median $E(B-V)$ of 0.56.

\subsection{Distance}

We determine the distances to the clusters using two methods:
spectroscopic parallax and ``main-sequence" fitting.  We discuss these
two approaches separately and then comment on their respective merits.

\subsubsection{Spectroscopic Parallax}
For each star with a spectral type we compute the distance modulus by
first dereddening the data using the intrinsic colors of FitzGerald
(1970), and then finding $V_o-M_V$.  We have adopted the spectral
type-$M_V$ calibration of Conti (1988) for our one O-type star and that
of Humphreys \& McElroy (1984) for everything else (B-type through M-type),
interpolating
values for spectral types not explicitly present in these tables when
needed.  For the stars with well defined spectral types we find a distance modulus
of 12.5$\pm$0.5 mag when we restrict ourselves to the objects within
the cluster cores.

This method works relatively well for very young ($<3$~Myr) clusters
(see Massey, Johnson, \& DeGioia-Eastwood 1995) where there is a fairly gradual change of $M_V$
with spectral type among the O-type stars (Conti 1983), about 1.4 mags
from O3~V to O9.5~V.   For a 10-20~Myr old cluster, the only stars
left on the main-sequence are B-type stars, and there $M_V$ changes
by 3 magnitudes over the spectral range B0~V to B8~V.  In addition,
the luminosity criteria for B-type stars are rather subtle compared to
that of the O-type stars;  for the former it depends upon the absolute
strengths of the Si lines (which are also temperature dependent), while
for O-type stars it depends primarily on He~II $\lambda 4686$ being in
emission or absorption.

We did examine the spectroscopic parallaxes as a function of luminosity
class. Although the median distance moduli for 
luminosity class V, III, and I stars
are about the same, in each case there is a large spread in the values
which we attribute to errors in placing the stars into the correct
luminosity class.  
Due to the subtlties involved, some misclassifications are
inevitable, and in addition membership issues may also come into play.
More significantly, stars which are really luminosity class IV and
hence follow their own (and presently undetermined) spectral type-$M_V$
calibration are placed at present into either luminosity class III or
class V.  The same holds for luminosity class II stars which often wind
up being called either class I or class III due to lack of observational
distinction between the classes.  

Using only the brightest stars
($V<10.5$) we find a distance modulus of $11.95\pm0.2$, in substantial
agreement with what we find below.  These brighter stars are among
the earliest types, yet include many supergiants whose
intrinsic luminosity has a large scatter.  However, 
the strength of the Si lines make spectral classification more
certain
in classifying hot giant and supergiant stars.   Therefore, spectral types
and luminosty classes will be better determined for these bright stars.

\subsubsection{Main-sequence Fitting}
We can also use ``main-sequence fitting'' to determine the average
distance of the lower-mass, unevolved stars in our sample.  Given the
approximate age of the clusters (13 Myr, as will be discussed in
section 3.4.3) this is not a trivial exercise and must be done using
post- and pre-main sequence evolutionary tracks combined.  Although many
renditions of the ``observed'' main sequence exist (e.g. Balona \& 
Shobbrook 1984; Balona \& Feast
1975; Morton \& Adams 1968; Blaauw, 1964), any relationship between M$_V$
and $(B-V)_o$ derived for stars in the solar vicinity necessarily contains
only the mean M$_V$ value characteristic of the typical age in the solar
neighborhood of stars with that $(B-V)_o$.  For example, the M$_V$'s at
the bluest values of $(B-V)_o$ represent ages of only 1-2 Myr while the
M$_V$'s around $(B-V)_o$=0 represent ages of a few hundred Myr and the
M$_V$'s around the $(B-V)_o$ color of the Sun represent ages of more than
a Gyr. In \hchi, plotting the de-reddened cluster data points against an
``observed'' main sequence results in the ``main sequence'' being too
blue/faint at the bluest colors (since the \hchi stars are evolved away
from the zero-age main sequence) and too bright/red at the redder colors
(since the \hchi stars are younger than the mean age of stars in the
solar vicinity and, hence, not yet far enough evolved from the empirical 
``zero-age'' main sequence to the position where stars having
the mean age of the solar neighborhood would lie).  Comparison to 
a theoretical "zero-age main sequence" involves similar concepts.    

``Main sequence fitting'' therefore must be done when trying to fit more evolved clusters using theoretical
isochrones. We use the 
solar-metallcity ($Z$ = 0.02) post-zero-age main sequence tracks and isochrones
of Schaller et al.\ (1992), which include convective overshoot and standard
mass-loss rates.  In addition, we use the pre-main sequence tracks
and isochrones from the same group, published in Bernasconi (1996).
We transform these tracks and isochrones from \mbol and \logt to V$_o$ and
$(B-V)_o$ using the same, though inverted, relationships that we use later
to transform our data from the observational V$_o$ and $(B-V)_o$ plane
to the theoretical \mbol and \logt plane (see Section~\ref{sec-HRD}).
Figure 6 is a CMD of our dereddened photometry where we
have used the isochrones to determine the distance.
As in Figure 4, blue and red
points represent stars which are spatially located within 7 arcmin from
the centers of \hPer  and  \ChiPer, respectively.

The ``main sequence fitting'' procedure is complicated by the fact
that in order to obtain a best fit distance modulus from theoretical 
isochrones an approximate age must be assumed.
We explored isochrones spanning a wide range in age (1-50~Myr) and found
that the isochrone shape matches the cluster data best for both post-
and pre-main sequence tracks in the 10-20~ Myr range.  
Using this
information, we find a best fitting distance modulus of 11.85$\pm$0.05 mag
corresponding to a distance of 2344$^{+55}_{-53}$ pc, where we have
estimated both the fit and uncertainty by eye in matching the
models to the data.   

Good agreement is seen between the post-zero-age main sequence turnoff at
$(B-V)_o \leq -0.22$ and pre-zero-age main sequence turn-on at $(B-V)_o
\geq 0.1$ using the Schaller et al.\ (1992) and Bernasconi (1996) calculations
{\it when transformed using our equations relating \logt with $(B-V)_o$
colors and bolometric corrections}.  Using transformations to the
color-magnitude plane supplied directly by the authors (which rely
upon the Schmidt-Kaler (1982) relationships) does not produce a match
between the theory and the data for any isochrone.  However, using our
transformation equations (derived primarily from stellar atmosphere
models) we find extremely good agreement between our dereddened data
and the stellar evolutionary isochrones.  As can be seen in 
Fig. 6, at an age of 10-15 Myr, we expect to see significant
contributions from the pre-main-sequence population at $M_v \sim$ 2.  
However, due the large amount of field star contamination is this region,
the extent of this effect in our sample is difficult to determine.

Fig. 6 effectively puts to rest any question
as to whether or not h and $\chi$ Per are at two different distances
rather than one.  This result is supported by similar conclusions
found by Keller et al.\ (2001). 

Why are the distance moduli derived from spectroscopic parallax (12.5
mag) and photometric parallax (11.85 mag) so different?  
The slightly evolved state of the main-sequence stars that dominate our 
spectroscopic sample should actually lead to our computing too {\it small}
a spectroscopic distance modulus rather than too great a number.  We believe
there is need for good recalibration of the spectral-type to $M_V$ relation
using
a variety of clusters and associations with good distance moduli determined
from spectroscopic parallax of O-type stars, as well as direct determinations
via modern trigonometric parallaxes.  

\subsection{The Hertzsprung-Russell Diagram}
\label{sec-HRD}

\subsubsection{Transformations}

The effective temperatures and bolometric corrections of our stars
were determined using photometry and spectral types, if available, or photometry
alone, otherwise, in order to place the stars in the HR diagram.
For those stars with spectral types we adopt the calibration
of Kilian (1992) for the early B dwarfs and giants, and that of
Humphreys \& McElroy (1983) for all other stars.  When spectral types
were not available, empirically derived relationships were used to
transform photometry to \logt and \mbol.
Effective temperatures were derived for bluest stars ($Q<-0.6$)
using the $Q$-\logt relationships given by Massey, Waterhouse, \&
DeGioia-Eastwood (2000), namely:

\noindent
$$\log T_{\rm eff}= -0.9894-22.7674\times Q-33.0964\times Q^2-16.19307\times Q^3 \phantom{XXXXXX} [\rm I]$$
$$\log T_{\rm eff}= 5.2618 + 3.4200\times Q+2.93489\times Q^2              \phantom{XXXXXXXXXXXXXXX} [\rm III]$$
$$\log T_{\rm eff}= 4.2622 + 0.6452\times Q+1.09174\times Q^2              \phantom{XXXXXXXXXXXXXXX} [\rm V]$$

\noindent
For stars that failed to meet this criterion,
we used empirical fits to a combination of ``observed" (Flower 1996) and
theoretical (Kurucz 1992) colors and effective temperatures.  The former
must be used with some caution as there is no reddening correction
for what are presumed to be nearby stars.  We found: 
$$\log T_{\rm
eff}=3.9889-0.7950\times(B-V)_o+2.1269\times(B-V)_o^2-3.9330\times(B-V)_o^3+$$
$$3.5860\times(B-V)_o^4-1.5531\times(B-V)_o^5+0.2544\times(B-V)_o^6$$.

The bolometric correction as a function of effective temperature is that
derived by Hillenbrand (1997) for dwarf stars but modified to account
for the presence of M supergiants in our sample by adopting the values
in Humphreys \& McElroy (1984).  Thus

$${\rm BC}=-8.58 + 8.4647 \times \log T_{\rm eff} - 1.6125 \times (\log T_{\rm eff})^2 \phantom{XXXXXXXX}[\log T_{\rm eff}>4.1] $$
$${\rm BC}=-312.90 + 161.466 \times \log T_{\rm eff} - 20.827 \times (\log T_{\rm eff})^2 \phantom{XXXXX}[4.1>T_{\rm eff}>3.83]$$
$${\rm BC}=-346.82 + 182.396 \times \log T_{\rm eff} - 23.981 \times (\log T_{\rm eff})^2 \phantom{XXXX}[3.83>T_{\rm eff}>3.55]$$
$${\rm BC}=-2854.91 + 1590.11 \times \log T_{\rm eff} - 221.51\times (\log T_{\rm eff})^2 \phantom{XXXXXXX}[3.55>T_{\rm eff}] $$

\subsubsection{The HR Diagram}

Figure 7 is the resulting HR diagram.  Post-zero-age main
sequence evolutionary tracks and isochrones are transformed as above
from the \logt and \mbol values calculated by Schaller et al.\ (1992).
All stars with MK spectral classifications of luminosity class I or
III, and stars earlier than B5 with luminosity class IV or V were
placed spectroscopically (larger, filled in circles) while most other
stars were placed photometrically (open circles).  The left panel
shows data for the entire imaging area while the right panel contains
only stars within 7 arcmin of the cluster nuclei.  No corrections for
field star contamination have been applied and, as was the case for
the color-magnitude diagrams (Fig. 5), the HR diagrams for the central regions
of the clusters contain significantly less field star contamination,
especially above the main sequence.   Note the presence of the
O-type star HD~14434 on the left panel of Fig. 7.  It
is highly discrepant in age, and, combined with its location outside
the cores of the clusters, we dismiss this star as field star.

From these HR diagrams we immediately see that the \hchi clusters are
slightly evolved from the zero-age main sequence and that the most
massive stars are only $\approx 20-30 \cal M_\odot$.  
The data extend down to about
$3 \cal M_\odot$ before field star contamination becomes substantial. 

\subsubsection{Stellar Ages and the Age Distribution} 

For finding ages, we use our dereddened CMD data 
(Fig. 6) with a grid
of isochrones computed at intervals of 0.1~Myr from 5-30~Myr.\footnote{We
use the CMD rather than the HRD data to determine ages in order to avoid
the quantization problem introduced by spectral types.  The ages are very
sensitive to $\log T_{\rm eff}$ (or $(B-V)_o$), and thus this quantization
would introduce a spurious age spread.  The spectral types {\it have}
been employed in the CMD in order to derive $E(B-V)$.  In a subsequent
section we {\it will} use the HRD to derive the mass function.  The masses
are primarily sensitive to an accurate determination of $M_{\rm bol}$,
which we expect to be better determined using the bolometric corrections 
determined from spectral types.} 
We restrict ourselves only to the most luminous stars ($M_V<-3$),
as it is only near and above the turn-off that there is good age information.
We filter out the obvious foreground contaminants
e.g., $(B-V)_o>-0.2$ for $-3>M_V>-5$.  We cannot use the RSGs for our
age determinations, unfortunately, since the evolutionary tracks do not 
actually extend that far to the red; we will note, though, that the location
of the RSGs in the CMD are consistent with the ages we derive were we to
extrapolate the isochrones.
For each of our clusters we find essentially identical ages:
12.8~Myr and  12.9~Myr for h and $\chi$ respectively.  The formal errors
of the mean on these determinations are 1~Myr, and the scatter
is $\sim 5$~Myr; the latter is dominated by observational errors at the
$<$0.01 mag level.

We do not find 
evidence for multiple distinct episodes of star formation despite the 
remarkable similarities between our dereddened CMD and Wildey's (1964).  
We believe 
the difference in interpretation occurs because Wildey in his original analysis
did not consider the possibility of field star contamination from G and K giants
seen to large distances through the Galaxy. It is clear from 
the right panel of Figure 7 that when just the cluster nuclei are considered
any apparent branching in the HR diagram is significantly diminished.  
We do find several high-mass stars 
with uncharacteristically young ages as compared to the rest of the cluster.  
However, in most cases these stars are either not in the central regions 
of the clusters or their spectroscopically derived distance is inconsistent 
with their being cluster members.   

Although our data are consistent with the h/$\chi$ stars having formed in
a single burst, we cannot rule out other scenarios.  For instance, if
the primary burst of star formation has occured at 13~Myr, with a smaller,
secondary burst at 
10~Myr, we would very likely not discern this in our CMD.
There would be few high mass stars, and 
the lower masss stars would be indistinguishable
from their 13~Myr counterparts.  

\subsubsection{Stellar Masses, the Mass Function, and Mass Segregation} 

Masses are inferred for individual stars by interpolating 
between the mass tracks on the HR diagram.  
By counting the number of stars found in each mass bin, we derive
the ``present day mass function" (PDMF). To the extent that star
formation may be coeval, this is equivalent to the initial mass function
(IMF) except for the depopulation of the highest mass bin.

In order to minimize the effect of field star contamination, 
PDMFs have been constructed only for the two regions within 
7 arcmin of the cluster cores.   
In addition, we exclude a few stars found redwards of the
main-sequence, and  presumed to be foreground
contaminants, by eliminating stars in the region 
constrained between
$M_{\rm bol}<-20.5\times \log T_{\rm eff} +82.5$ and
$M_{\rm bol}>-5$.
We used a lower mass cutoff of ~4 $\cal M_\odot$ below
which 
field and pre-main
sequence star contamination dominate.  At the high-mass end, we expect
that evolution through the supernova phase
will have depleted stars above $\sim 15-20 \cal M_\odot$,
and so we have used only the mass bins below this to compute the
slope of the IMF.  
We combine all of the higher-mass stars into one mass bin.
Following Scalo (1986),
we define the quantity $\xi$ as the number of stars per mass bin
divided by the difference in the base-ten logarithm of the upper and lower
bin masses,  and also by
the surface area in kpc.  The run of $\log \xi$ with
log mass then provides the slope, $\Gamma$, of the IMF/PDMF.  Values for
the number of stars and for $\xi$ are given in Table~3.  

Figure 8 shows PDMFs in the
4-16 $\cal M_\odot$ range for stars within 7 arcmin of the cluster
centers.
Error bars are based on $\pm$$\sqrt{N}$
statistics.
We obtain values of $\Gamma=-1.36\pm0.20$ for \hPer and $\Gamma=-1.25\pm0.23$
for \ChiPer.  Within the errors of our fits,
both slopes 
are in good agreement with each other and also with the
Salpeter value of $\Gamma=-1.35$.
This result can be compared with what is known of the IMF in other young
OB associations and clusters, where a weighted average yields
$\Gamma=-1.1\pm0.1$ for the Milky Way $\Gamma=-1.3\pm 0.1$
for the LMC/SMC (Massey 1998b).  Thus, an IMF slope of $\Gamma=-1.3\pm0.2$
for h and $\chi$ is in no way unusual.  

Based on extrapolation of the measured PDMFs to
120 $\cal M_\odot$, we estimate that  $\sim$40 supernovae have occured
in the past in the central regions of the \hchi clusters.
Assuming a constant mass function from 1-120 $\cal M_\odot$, we 
can estimate the total stellar mass within
each of the cluster centers down to 1 $\cal M_\odot$.  We find
values of 3700 $\cal M_\odot$ and 2800 $\cal M_\odot$ for \hPer and
\ChiPer,
respectively.   
This is about 8-10 times that of the mass in
$>$1 $\cal M_\odot$ stars
in the younger Orion Nebula cluster ($\sim 450 \cal M_\odot$)
or the older Pleiades ($\sim 320 \cal M_\odot$). 
For comparison, a ``supercluster" like R136
in the LMC has a mass of roughly $3-4 \times 10^{4} \cal M_\odot$
in $>$1 $\cal M_\odot$ stars (Hunter et al.\ 1996), about
a factor of 10 greater than either h or $\chi$ and a factor of almost 100
greater than Orion or the Pleaides.

In Figure 9 we explore the evidence for concentration and
mass segregation in the two clusters.  In doing so, we consider
only those stars satisfying our criteria for inclusion in the PDMF.  
In viewing these panels it should be noted that the 2$\sigma$ surface density 
contour in the spatial distribution of stars
occurs at radii of $\sim$7 arcmin for both \hPer and \ChiPer.
The top and middle panels of Fig. 9 show that
inside of 7 arcmin, both the mass surface density and the number surface 
density begin to rise noticeably above the field star surface density, and then
steepen considerably at $\sim$3 arcmin.  The increase in density 
at smaller cluster radii is evidence of higher central concentration.

The histograms of the total mass/pc$^2$ as a function of radial distance 
(top panels of Fig. 9) show that \hPer is about twice 
as dense at its core compared to \ChiPer.  This occurs both because \hPer has 
$\sim$25\% more stars at its center (as can be seen in the middle panels 
of Fig. 9) and because it contains several high mass 
($>30 \cal M_\odot$) B supergiants. 
However, the density profile of \hPer falls off 
more rapidly than that of \ChiPer and the two clusters are roughly equivalent 
in mass density at a radius of $\sim$3 arcmin.

The bottom panels of Fig. 9 show the average mass
as a function of radial distance from the cluster centers. For \hPer,
we find a significant gradient inside of $\sim$7 arcmin in the
mean mass {\it vs.} radial distance, suggestive of mass segregation.
The data for \ChiPer is less convincing, yet we still find the
mean stellar mass to be higher by $\sim1.5-2\sigma$ within the central 1 arcmin.
This phenomenon has been claimed with varying degrees of strength in other 
open clusters in the Galaxy (e.g. the Orion Nebula Cluster; 
Hillenbrand \& Hartmann 1998 and references therein) and in the 
Magellenic Clouds (e.g. R136; Hunter et al 1995, and NGC 1805 and NGC 1818; 
deGrijs et al 2002).
However, unlike their younger counterparts, the mean mass gradient in \hchi
may not be primordial, i.e. associated with the formation of the clusters.
Assuming a velocity dispersion of $\sigma_v \approx$3 km/s and a 7 arcmin
(4.79 pc) cluster radius, we estimate a crossing time of $\sim$1.56 Myr 
for each of the cluster nuclei.  Given that the clusters are
$\sim$13 Myr old, the age/$t_{cross}\approx$8 and hence dynamical
relaxation may indeed play some part in the observed mass segregation.

\section{Discussion}

\subsection{Comments on the Uniqueness of \hchi}

The \hchi clusters are separated by about 30 arcmin on the sky, equal to 20 pc,
and are located $\sim$3.5$^\circ$ or 140~pc out of the plane of the Galaxy. 
They are thus similar to but larger and more massive than the younger, closer
Orion Nebula cluster and NGC 2024 pair which are separated by $\sim$32~pc and 
located $\sim$150 pc out of the plane, or the IC 348 and NGC 1333 pair 
$\sim$21 pc from each other and $\sim$122~pc from the plane. As noted above, however,
the massive star content of \hchi is more than an order of 
magnitude higher it is in these regions.  Other suggested 
coeval double cluster systems include the older SL 538 / SL 537,    
SL 353 / SL 349, SL 387 / SL 385, NGC 1971 / NGC 1972, and NGC 1850 pairs 
(e.g. Dieball \& Grebel 1998, 2000a, 2000b and references therein)
all in the LMC (see Bhatia \& Hatzidimitriou 1988 and 
Hatzidimitriou \& Bhatia 1989 for other LMC and SMC candidates), 
and the young NGC 206 (van den Bergh, 1966; see also Massey, Armandroff,
\& Conti 1986) in M31.  The range 
in scale of ``double-cluster" formation may extend, therefore, from clusters 
of individual size ranging from a few pc up to a few hundred pc in diameter.
In the younger of these double clusters, for example
the ONC / NGC~2024 pair, the stellar and cluster dynamics are still dominated 
by molecular gas and the clusters are at best only marginally/loosely bound 
once the gas dissipates, unlike \hchi which have survived as bound clusters 
for $\sim$10 Myr after gas dissipation.  At present, however, 
kinematic studies of the \hchi cluster motions relative to one another are needed
in order to decipher whether the \hchi clusters are a true binary system, 
or merely reflective of synchronized star formation on larger size scales.

Despite the above suggestion that double-cluster formation may be fairly 
common, we now argue that \hchi are nearly unique.  They are remarkably 
similar clusters insofar as we find their distances, reddenings,
ages, IMF slopes, and physical sizes to be indistinguishable.  The stellar
density of H Per, however, is a factor of 2 higher than that of \ChiPer and
its total mass about 1/3 more.
Independent of whether this single difference is considered or ignored,
the \hchi system evokes the word ``unique" when considered in the context
of the Galaxy.  The system is, after all, commonly known as 
{\it the} double cluster.  While the LMC may contain a high proportion of 
double clusters that are coeval, these systems show a wide range in
total mass ratio and size ratio (see, e.g. Leon, Bergond, \& Vallenari (1998)). 

\subsection{Comments on the Relationship Between \hchi and Per OB1}

The \hchi clusters are often described as the core of the Per OB1 association, 
located in the Perseus spiral arm at a distance of $\sim$2.3 kpc 
(Humphreys 1978; Ruprecht 1966).  A similar relationship between clusters 
and associations may hold in other cases, such as the pair of open clusters 
IC 1805/IC 1848 and Cas OB6, seen in projection only 5 degrees 
from h/$\chi$.  Garmany \& Stencel (1992) question the physical relation between 
\hchi and the Per OB1 association, other than being located along the 
same line of sight and in the same spiral arm which is nearly perpendicular 
to our line of sight in the direction $l\approx 135^\circ$.  That we see
\hchi projected in a field star distribution that is, to within a factor
of 30\%, at the same distance as the clusters complicates discussion of
the cluster / OB association relationship.  This discussion is further 
complicated by the similarity in age between the Per OB1 ``field'' population
and the \hchi clusters.  

Per OB1 is particularly notable for
containing the largest number of red supergiants (RSGs) among the 
associations whose high-mass members were catalogued (e.g.
Blanco 1955;  Humphreys 1970;
Garmany \& Stencel 1992; see also Bidelman 1947b), 
as well as a substantial number of A- and B-type supergiants (Bidelman 1943).  RSGs are 
visible only for a narrow range of ages between 10-25 Myr at the completeness
limit of our photometry and considering the distance to the Perseus spiral arm.
Thus it is difficult to reconcile whether the red supergiants at large projected
distances from the \hchi cluster (black points at \logt = 3.5 in the left
panel of Fig. 7) are part of the ``field'' or the
result of past ejection from the \hchi cluster core regions.
Ejection of massive stars from a dense cluster can occur for particular 
binary and system orbital parameter combinations, but requires that the cluster
is mass segregated at very young ages, e.g. at or before the time of gas 
expulsion.  (Kroupa, 2002). Populating the entirety of the Per OB association 
with stars ejected from the centers of \hchi is unlikely, though the effect 
may be as large as 10-30\% by the present cluster ages.  The double nature 
of the cluster may also be important for stellar dynamics considerations.

Along the main sequence of the HR diagram we find 
reasonable agreement between the cluster and ``field'' populations
again due to the similarity in distance and age of the massive star
population. But this does not prove physical association between
the clusters and the ``field'' or association.  Even kinematic information 
would be of limited use in this debate, given the magnitude of the effect
compared to achivable errors.

\subsection{Future Work}

At an age of 13~Myr, the \hchi clusters
occupy a particularly interesting age range for investigations of
circumstellar disk dissipation and of stellar angular momentum evolution.
The evolutionary paths of these phenomena are very poorly understood inbetween
the age ranges of well-studied star-forming regions ($<$1-3 Myr) and the nearest open clusters (IC 2602 and IC 2391 at 50 Myr, $\alpha$ Per at 80 Myr, and
the Pleiades at 120 Myr). Despite the larger distance relative to
some of these other well-studied open clusters, investigations of the
lower-mass ($<3 M_\odot$) stellar content of \hchi is therefore of great
interest.  Substantial field star contamination will complicate this issue
and require selection techniques such as x-ray or H$\alpha$ emission,
or photometric variability, to separate young active candidate cluster members from
the Galactic plane foreground/background in a photometric survey.

\section{Summary}

We have studied the h and $\chi$ ``double cluster" using modern
imaging and spectroscopic techniques.   We find that the two
clusters have indistinguishable reddenings ($E(B-V)=0.56$)
and distances [$(m-M)_o=11.85$], values consistent with those cited in
the literature. 
Especially impressive is that 
these conclusions are identical to those of Wildey (1964),
whose data we find holds up extremely well against the current analysis.

Where we differ with previous studies such as Wildey's (1964) is in our recognition of
the significant effect that field star contamination has on the determination
of cluster ages.  Inclusion of foreground younger stars and GK giants
can easily lead to apparent branching in the HR Diagram which has been
misinterpreted in the past as an age spread.  We find mean ages of 12.8 Myr
for each of the two clusters and no evidence for multiple epochs of star
formation.  

The present day mass function yields a slope consistent with that found in
other well-studied Galactic OB associations and 
clusters ($\Gamma\sim -1.1\pm0.1$, see Massey 1998b), and is essentially
Salpeter ($\Gamma=-1.35$).
In addition, we do find some evidence of mass segregation.  The total masses
are 3700 $\cal M_\odot$ for \hPer and 2800 $\cal M_\odot$ for $\chi$ Per,
for stars with $>1 \cal M_\odot$.

\section{Acknowledgements}
Our interest in h and $\chi$ Per traces back to several conversations
with Stephen Strom, who remarked at least once how hard it was to
understand a 50~Myr age spread, lamenting that ``If we don't understand
star formation in h and $\chi$ Per, then where do we understand it?"
We hope that we have partially addressed this concern.
It is a pleasure to thank Michael Meyer and John Carpenter
for work they did in measuring coordinates of the h and $\chi$ stars
which we used for spectroscopy prior to our CCD imaging efforts.
We also acknowledge help and advice from
George Jacoby in obtaining the Mosaic data.
CLS became involved in this project as a Research Experiences for Undergraduates
participant in 1999, and her efforts were
supported first by the National Science Foundation (NSF) under grant 99-88007
to Northern Arizona Unversity and more recently by an NSF Graduate Research Fellowship. 
We dedicate this paper to the memory of Bob Wildey, whose PhD
thesis on the subject of h/$\chi$ Per should be required reading
for all students of Galactic astronomy.

\clearpage

\begin{figure}
\epsscale{0.55}
\plotone{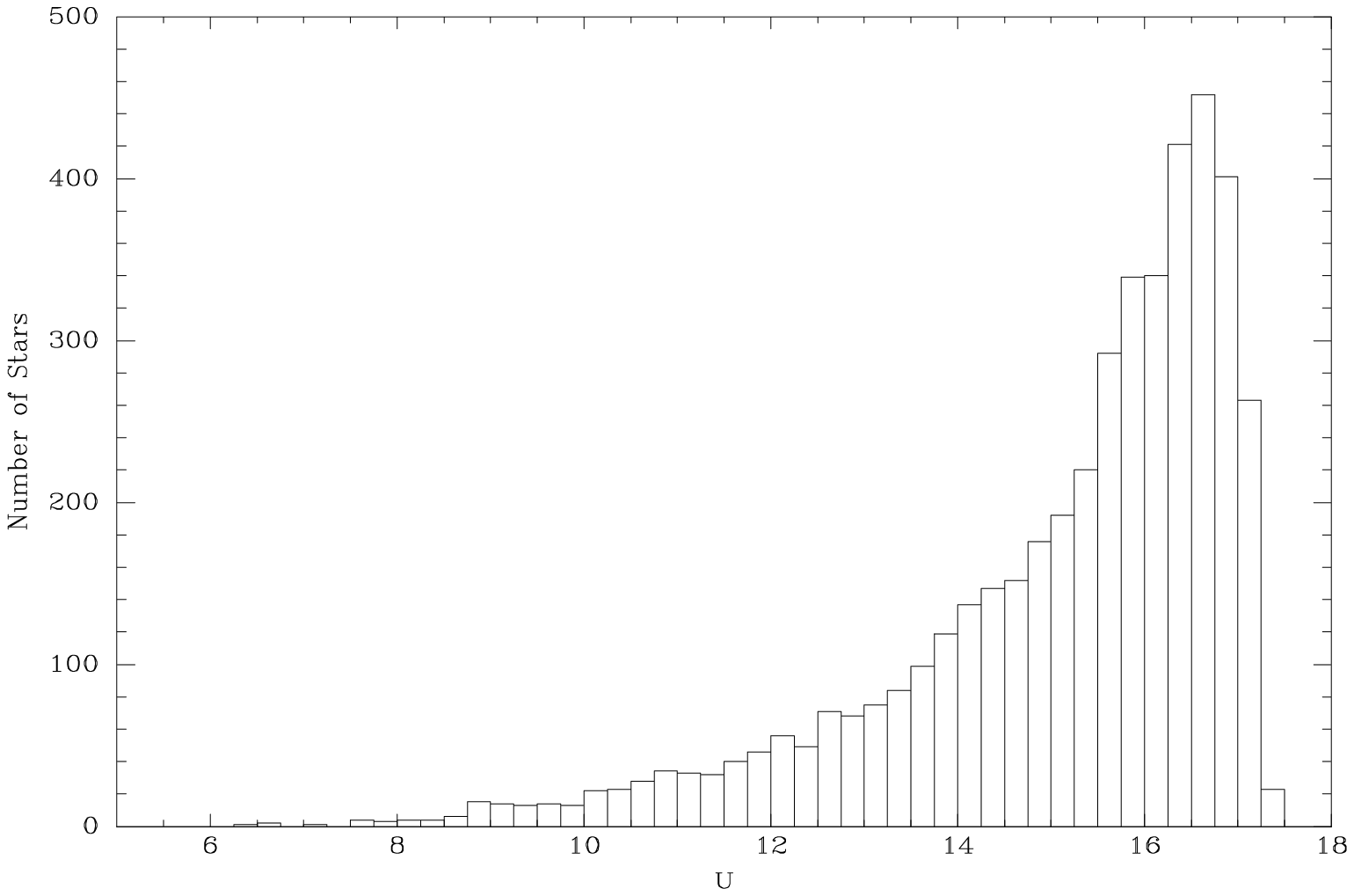}
\plotone{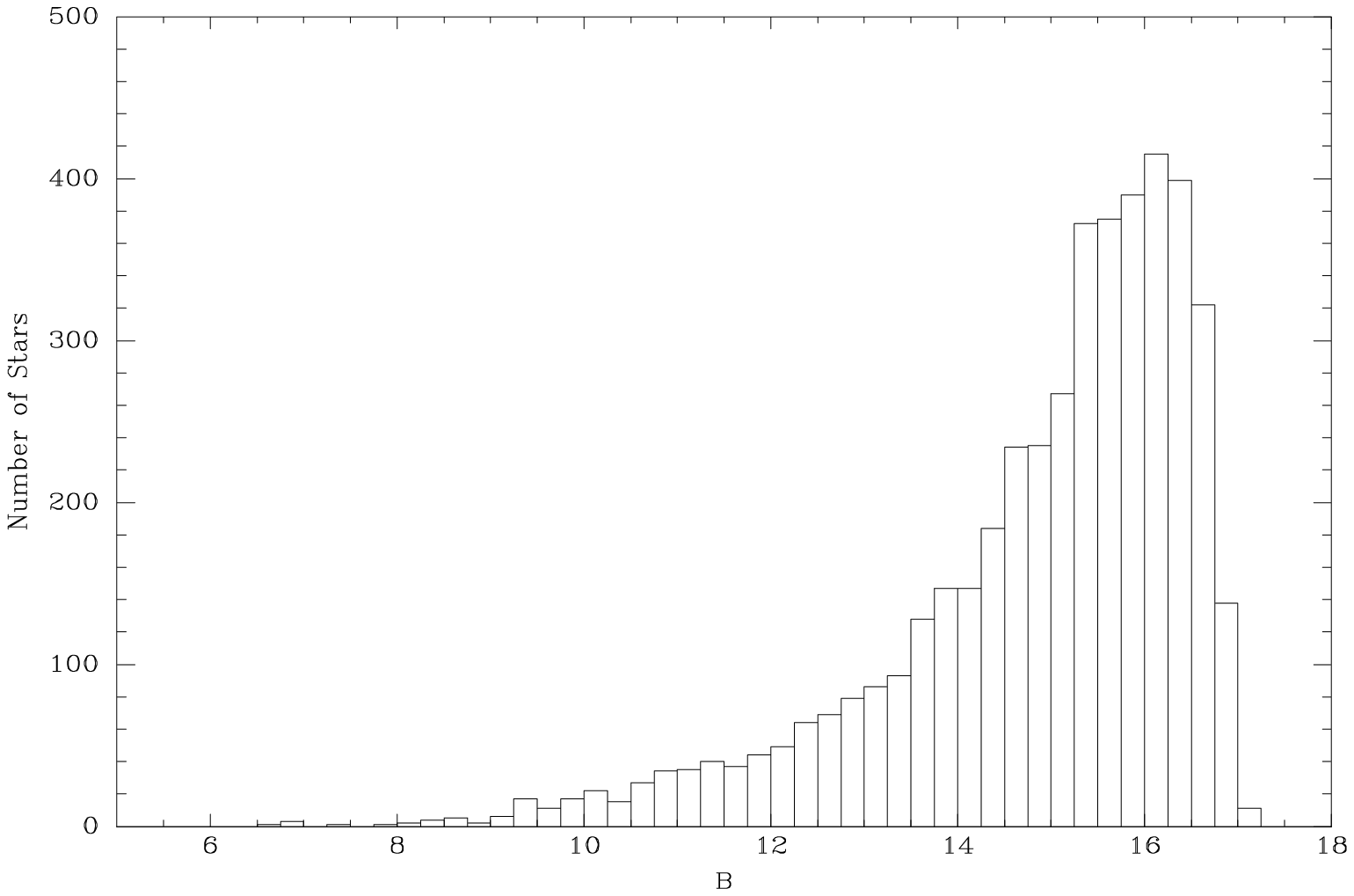}
\plotone{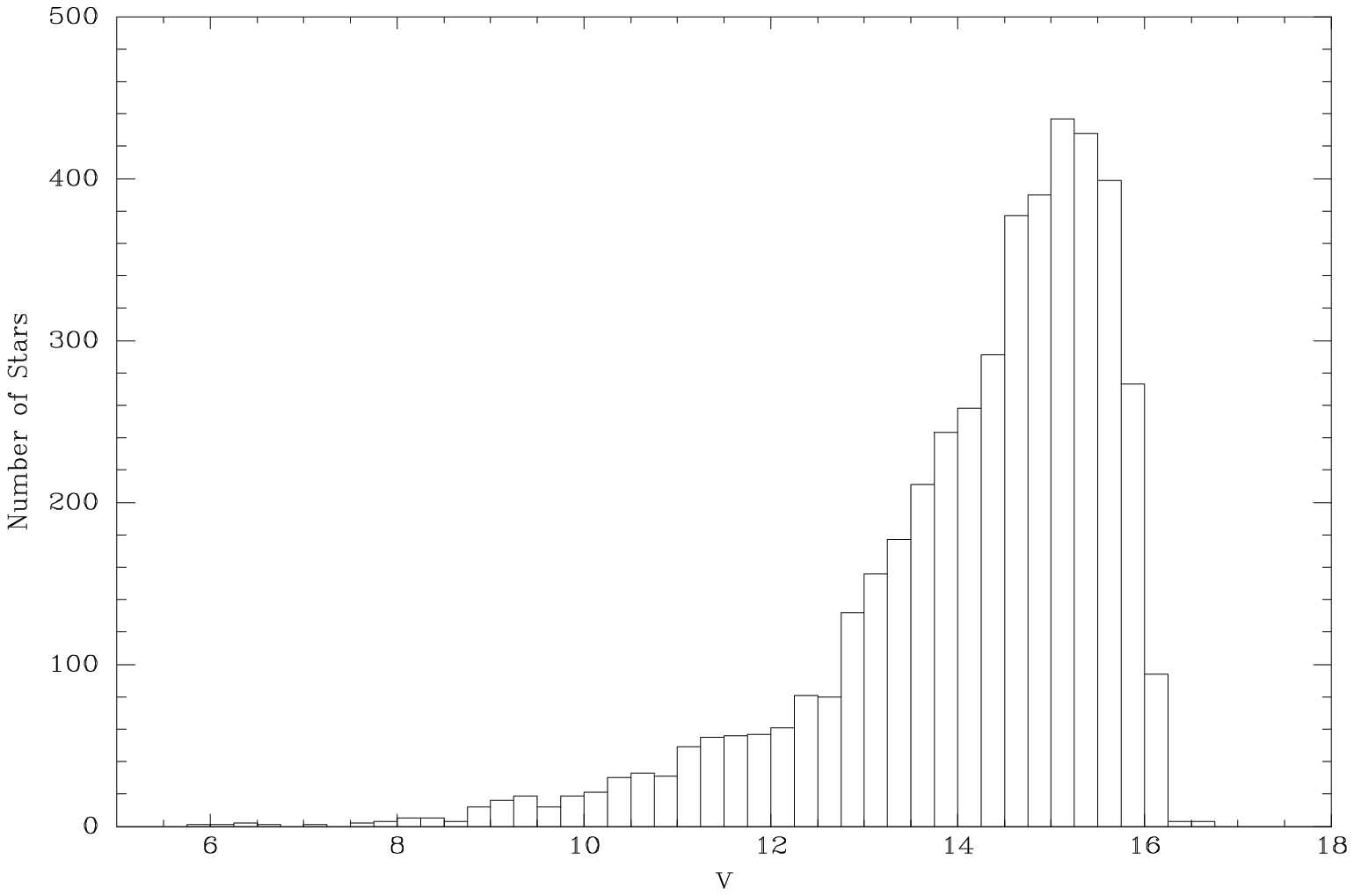}
\caption{For each of our three filters, 
we use the histogram of the number of stars 
as a function of
magnitude to estimate our completeness.}
\end{figure}

\begin{figure}
\epsscale{0.35}
\plotone{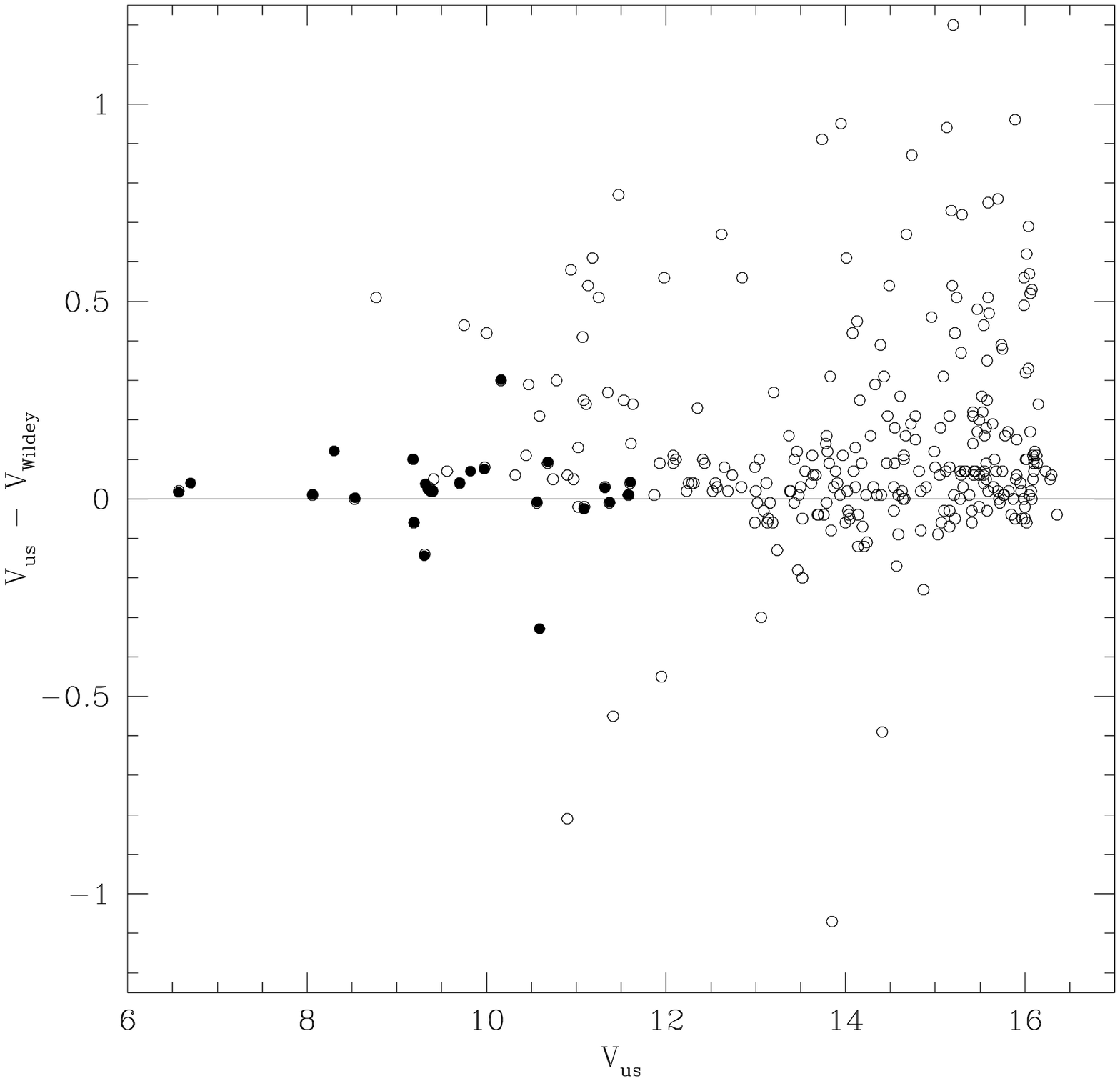}
\plotone{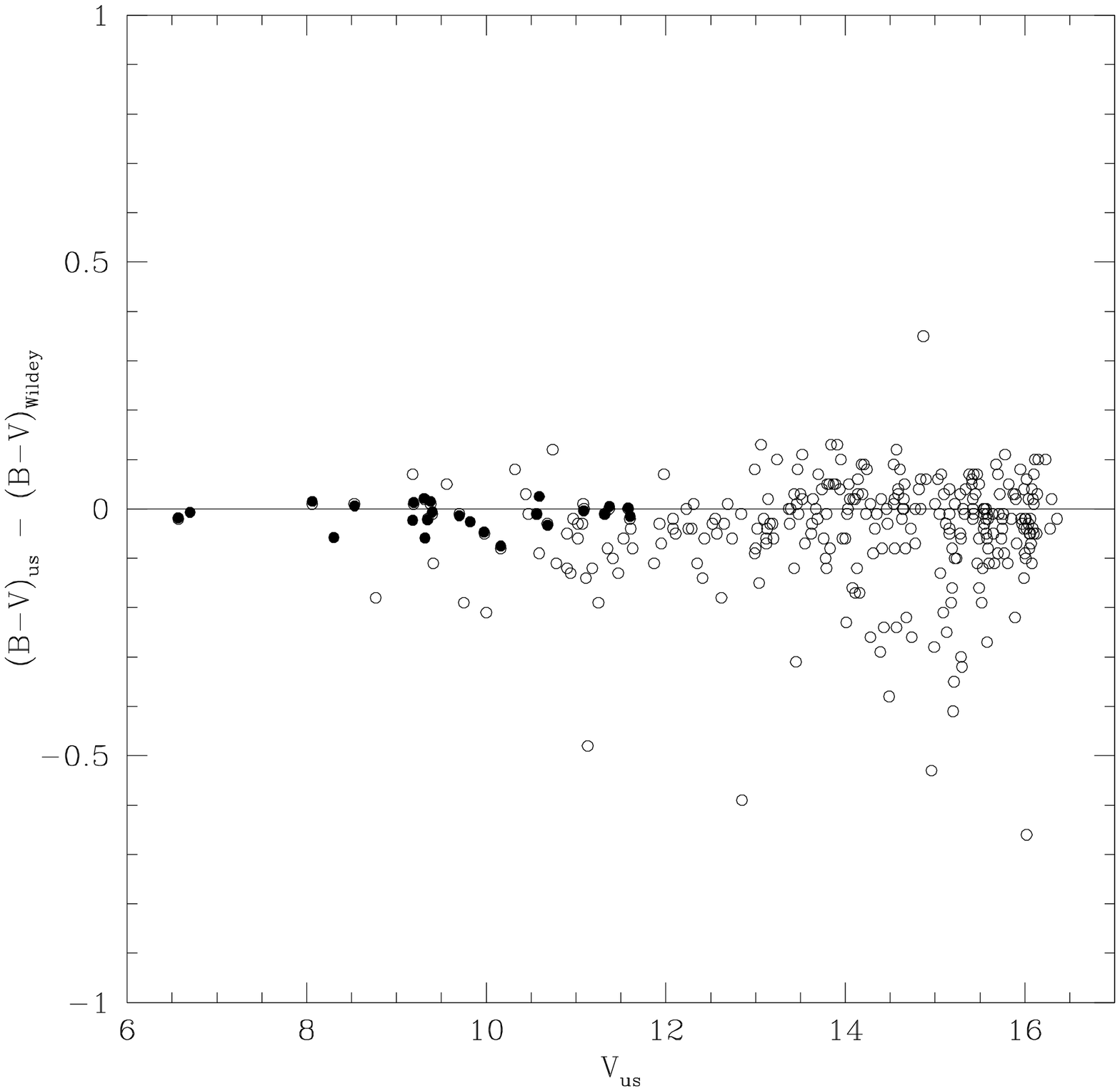}
\plotone{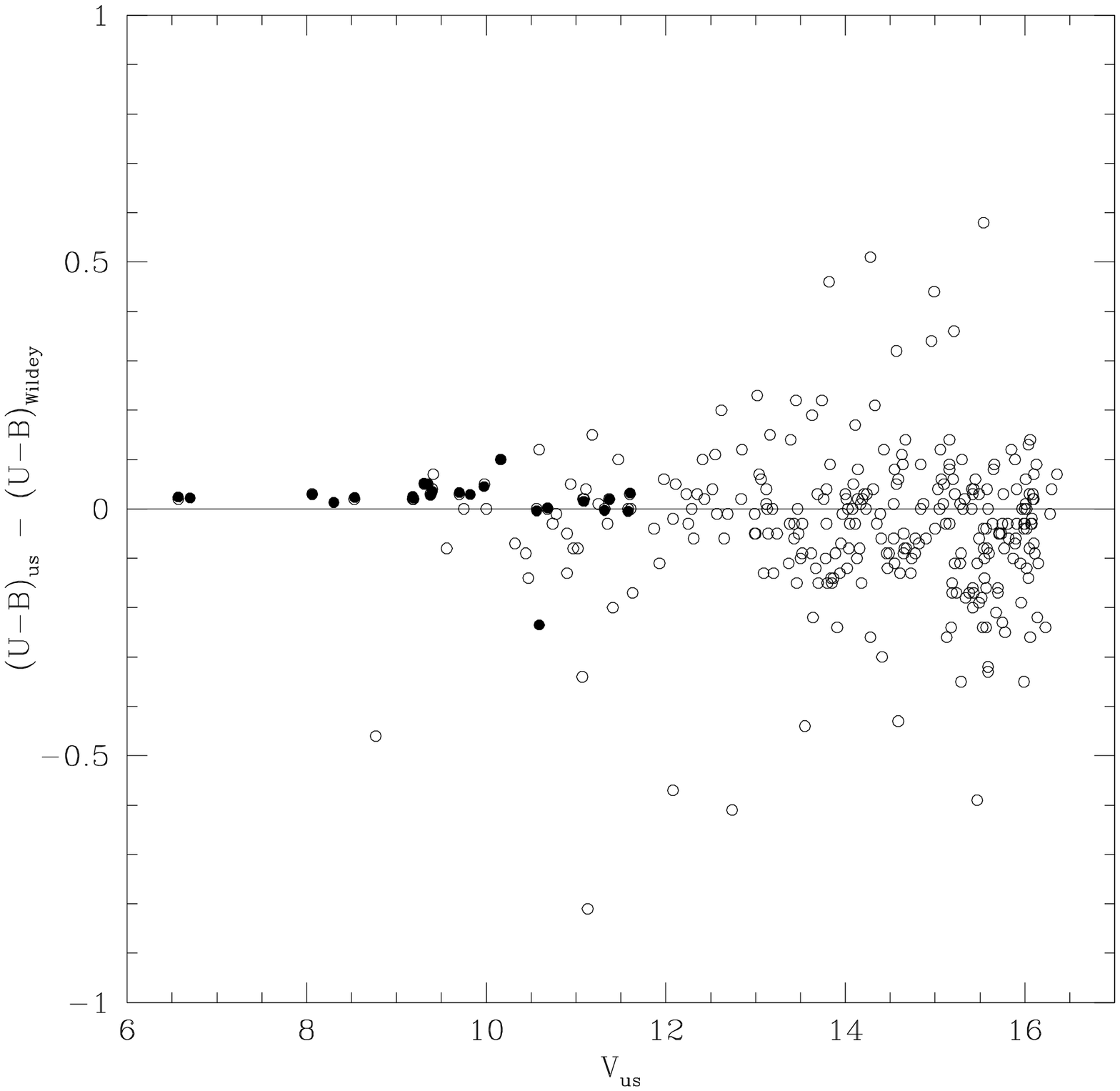}
\plotone{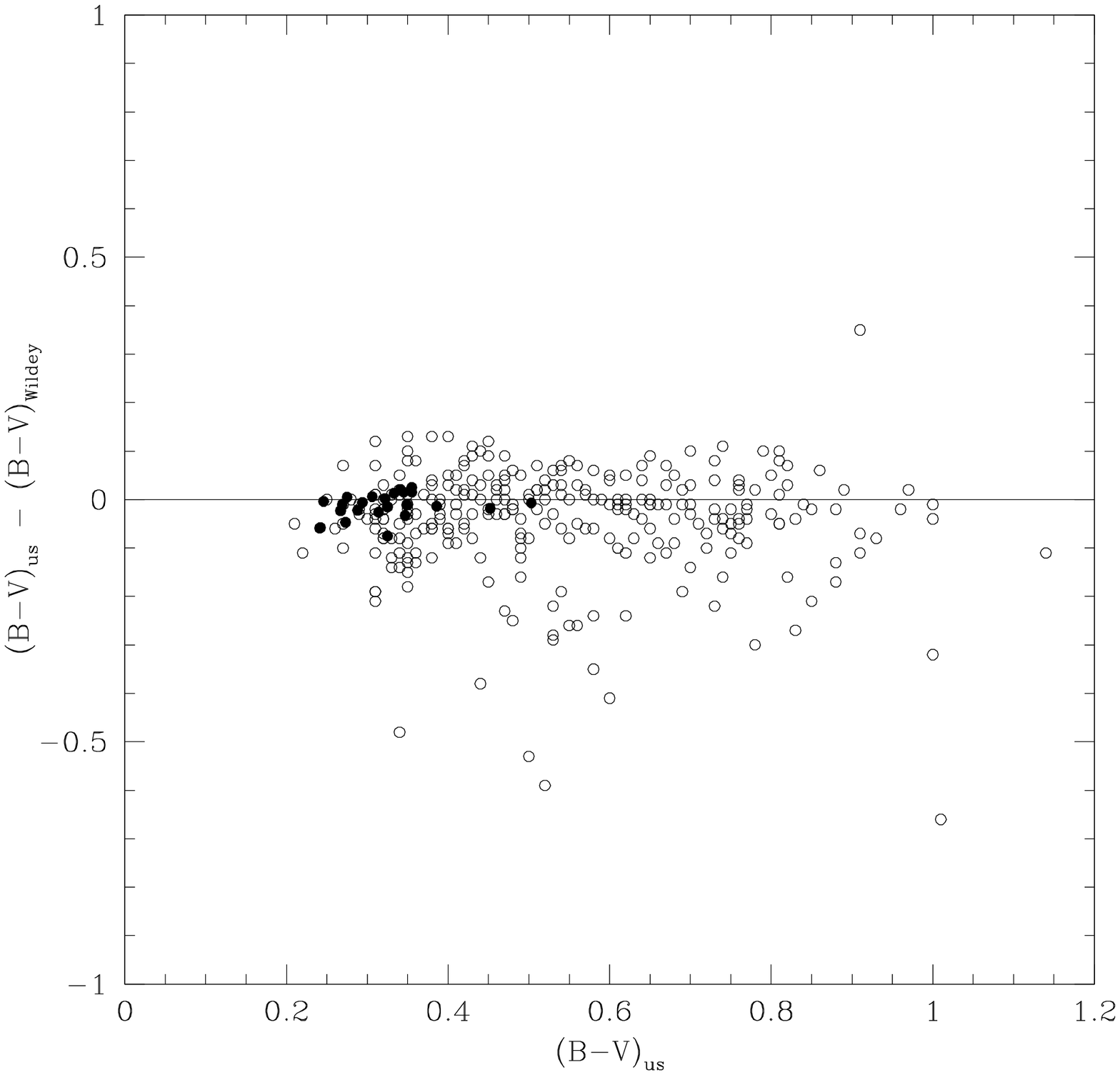}
\plotone{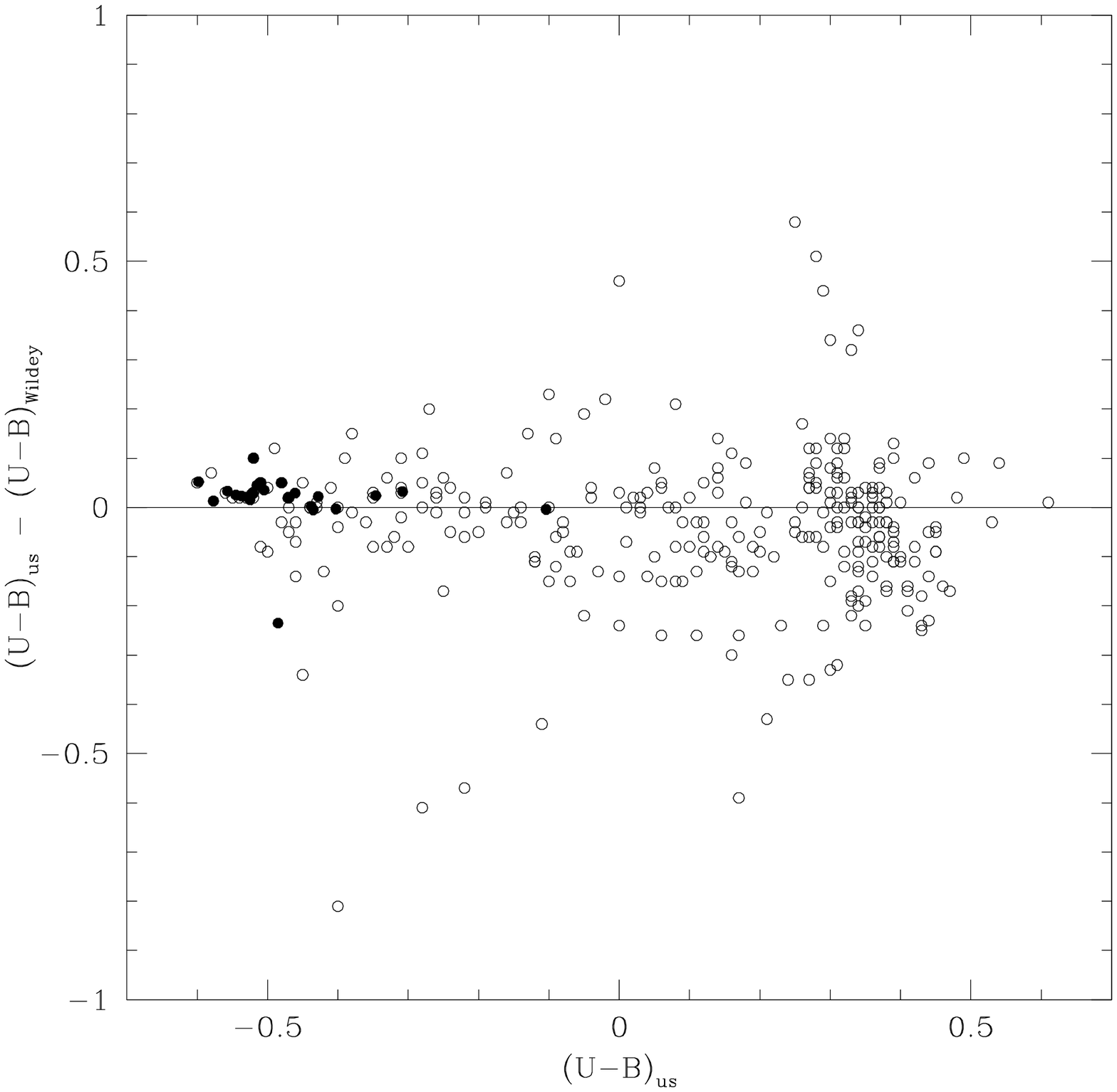}
\caption{Our CCD photometry is found to agree well 
with that of Wildey's (1964) data.  Open and closed circles are a comparison to
his photographic and photoelectric photometry, respectively.}
\end{figure}

\begin{figure}
\epsscale{0.8}
\plotone{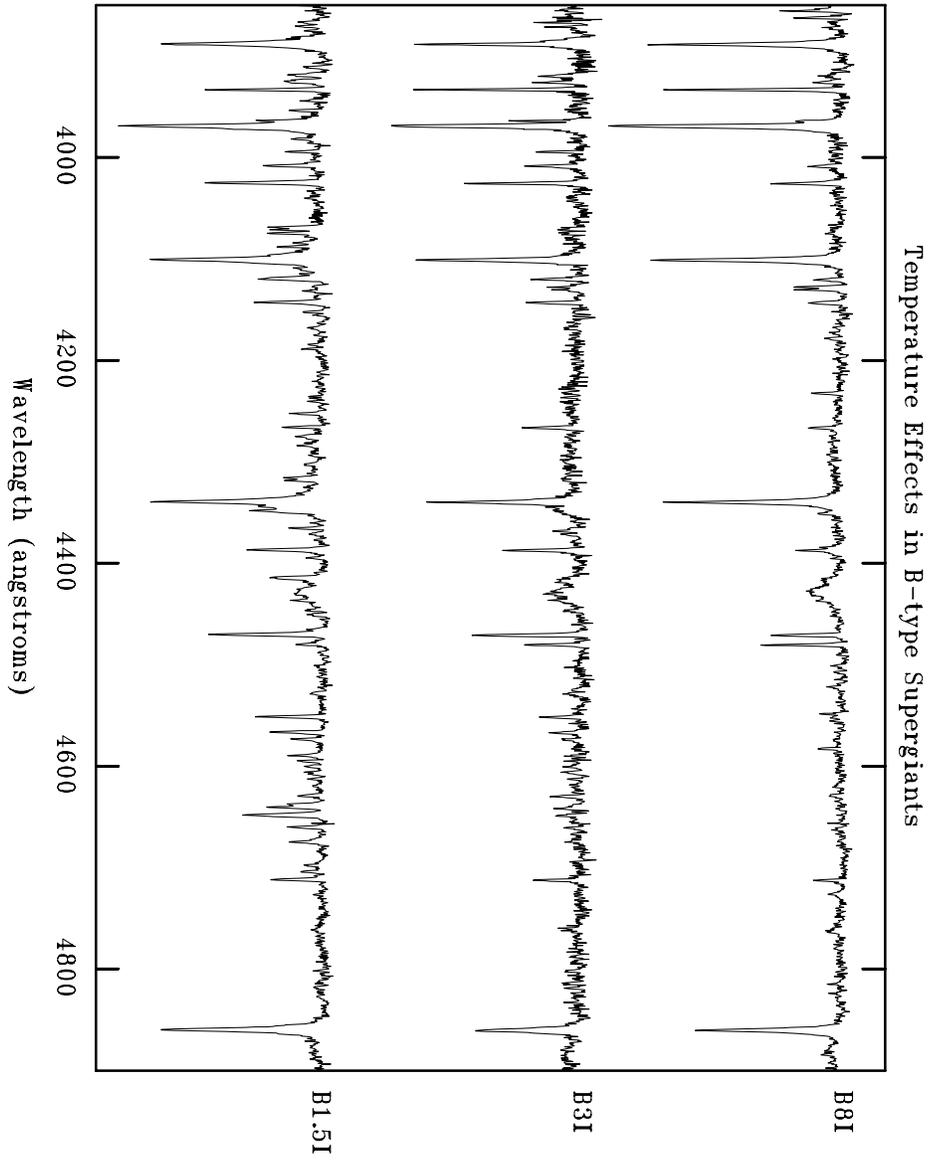}
\caption{Temperature effects in B-type supergiants.  The primary indicator
is the ratio of MgII 4481 to HeI 4471.}   
\end{figure}

\begin{figure}
\epsscale{0.75}
\caption{THIS FIGURE AVAILABLE IN .GIF FORMAT
On the left we show the 
uncorrected CMD for all of the stars in our sample
(upper panel), along with their spatial distribution (lower panel).  We
have overlayed colored symbols on the stars for which we have spectral types.
These may be compared to the diagams on the right, where
we have included only stars within 7~arcmin of the nuclei of the
two clusters. Stars near the center of h Per are indicated in blue,
while stars near the center of Chi Per are indicated in red.}
\end{figure}

\begin{figure}
\epsscale{0.8}
\caption{THIS FIGURE AVAILABLE IN .GIF FORMAT
The top panels show the
full CMDs plotted as contoured Hess diagrams, where we have overlayed color
contours to indicate the field-star contamination.  In the bottom panels we
have removed the field-star contamination and smoothed the data.
}
\end{figure}

\begin{figure}
\epsscale{0.8}
\caption{THIS FIGURE AVAILABLE IN .GIF FORMAT
The dereddened CMD for stars within 7 arcminutes of the center
of h (blue) and $\chi$ (red) Per are shown now with the ZAMS and post-main-sequence 
isochrones of 10 and
20 Myr indicated.  Corresponding 10 and 20 Myr pre-main-sequence
isochrones are shown as dashed lines.  The black points represent the rest of the stars in
our full $ 0.98^\circ \times 0.98^\circ$ field.}
\end{figure}

\begin{figure}
\epsscale{0.8}
\caption{THIS FIGURE AVAILABLE IN .GIF FORMAT
The HR diagram of h and $\chi$ Per are shown.  On the left
we show all of the data, with filled circles showing the data placed
by means of spectroscopy, and open circles being the data for which
have only photometry.  On the right we show only the stars within
a 7 arcminute radius of the center of h (blue dots) and $\chi$ (red dots).}
\end{figure}

\begin{figure}
\epsscale{0.8}
\plotone{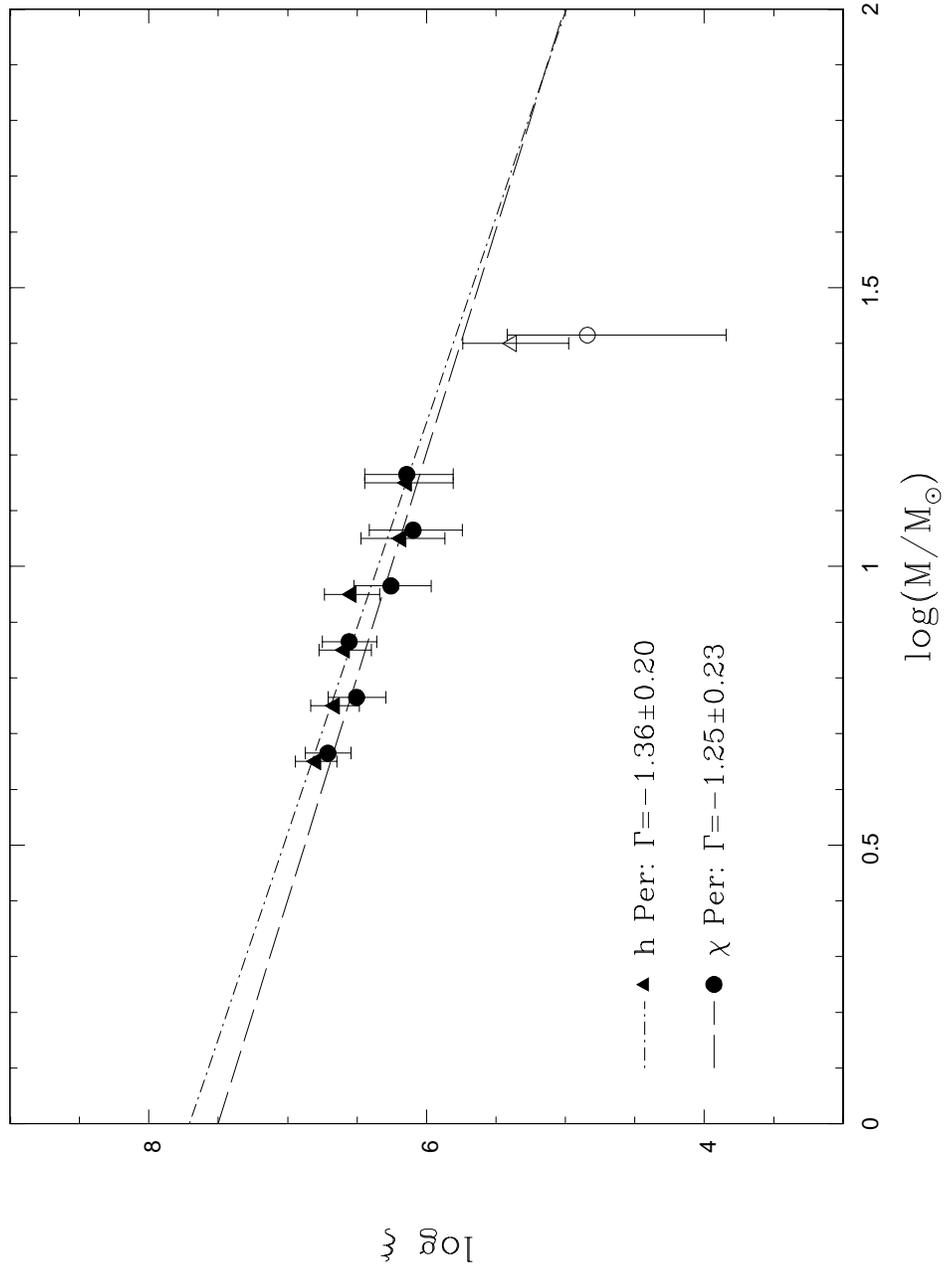}
\caption{The initial mass function is shown for the two clusters. Open symbols
indicate an incomplete bin.}
\end{figure}

\begin{figure}
\epsscale{0.8}
\plotone{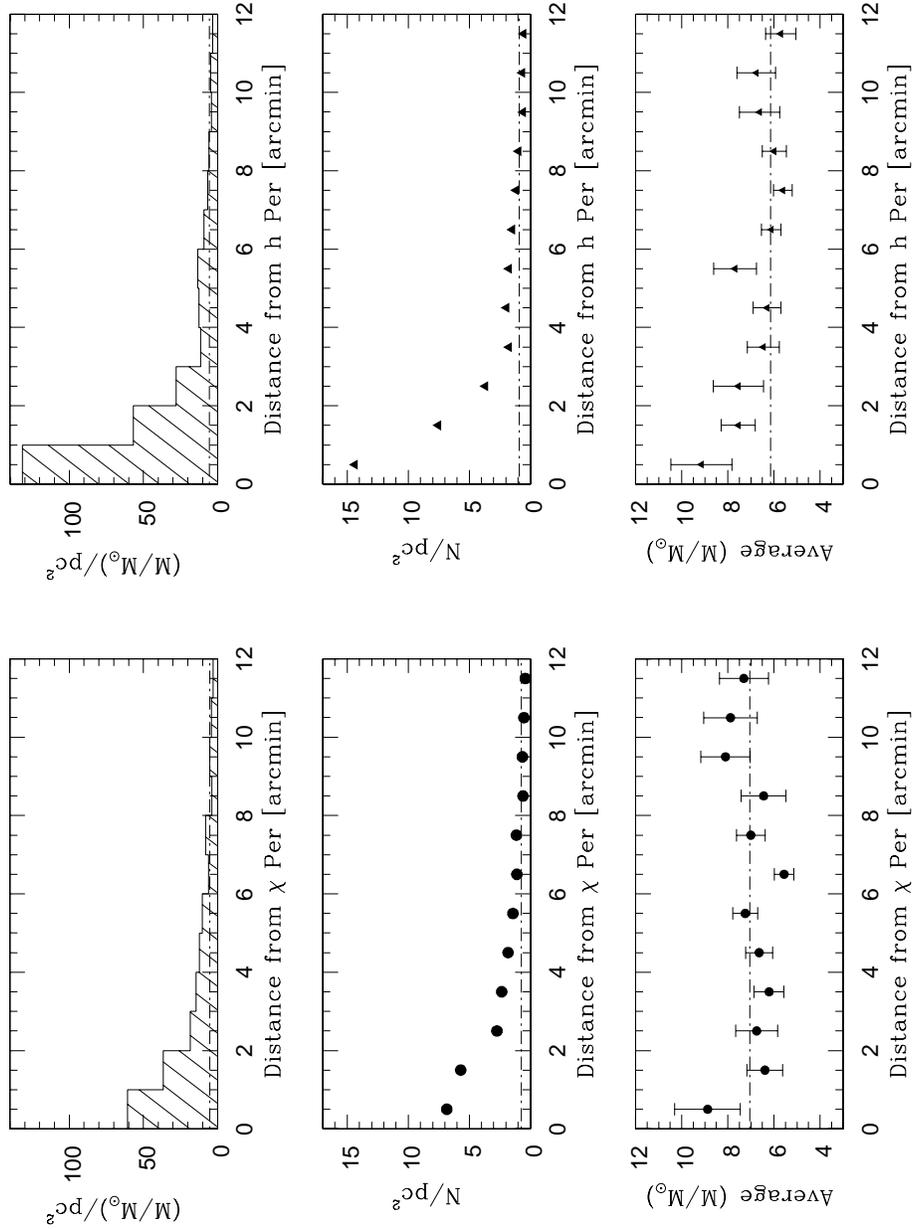}
\caption{The total mass per unit area (top panels), the
number of stars per unit area (middle panels),
and the
average stellar mass per unit area (bottom panels) are shown
as a function of radial 
distance for h Per (left) and $\chi$ Per (right). The data have 
been binned in 1 arcmin rings from the respective cluster centers.  Horizontal 
dashed lines indicate the average of the values from 6-12 arcmin.  Within the 
2$\sigma$ surface density enhancements (r = 7 arcmin), there is clear evidence 
in both h and $\chi$ Per for central concentration within 3 arcmin
(top and middle panels), and also some evidence for mass segregation within
1-2 arcmin (bottom panels).
}
\end{figure}

\clearpage
\begin{deluxetable}{cccccccccccc}
\tabletypesize{\scriptsize}
\tablecolumns{12}
\tablenum{1}
\tablecaption{Observational Data\tablenotemark{a}}
\tablehead{
  \colhead{} &
  \colhead{} &
  \colhead{} &
  \colhead{} &
  \colhead{} &
  \colhead{} &
  \colhead{} &
  \colhead{} &
  \colhead{} &
  \multicolumn{2}{c}{Spectral Type} & 
  \colhead{}  \\
  \cline{10-11} \\
  \colhead{ID} & 
  \colhead{$\alpha_{2000}$} &
  \colhead{$\delta_{2000}$} &
  \colhead{$V$} &
  \colhead{$(B-V)$} &
  \colhead{$(U-B)$} &
  \colhead{$d_h$\tablenotemark{b}} &
  \colhead{$d_\chi$\tablenotemark{b}} &
  \colhead{Assoc} &
  \colhead{Adopted\tablenotemark{c}} &
  \colhead{Literature\tablenotemark{d}} & 
  \colhead{Alt. ID\tablenotemark{e}}  
}
\startdata
1 & 2:18:04.44 & 57:30:58.9 & 6.023 & 1.097 &  0.832 & 0.389 & 0.681 & far & G7III* & G7III (4) & HD13994 \\ 
2 & 2:21:55.32 & 57:14:34.6 & 6.480 & 0.502 & -0.050 & 0.403 & 0.129 & far & A0I & A1Ia (2) & HD14433 \\ 
3 & 2:19:04.37 & 57:08:08.4 & 6.567 & 0.452 & -0.346 & 0.016 & 0.425 & h & B3I & B3Ia (1) & HD14134  \\ 
4 & 2:19:13.86 & 57:10:09.8 & 6.700 & 0.503 & -0.428 & 0.035 & 0.405 & h & B3I & B2Ia (1) & HD14143  \\ 
5 & 2:23:00.35 & 57:23:13.5 & 6.977 & 0.707 & -0.249 & 0.587 & 0.288 & far & B8I & B8Ia (2,3) & HD14542 \\
...&...&...&...&...&...&...&...&...&...&...&...\\ 
4528 & 2:19:35.40 & 57:18:29.9 & 16.906 & 0.160 & -0.609 & 0.176 & 0.400 & far & \nodata & \nodata & \nodata \\ 
\enddata

\tablenotetext{a}{The complete version of this table will be published in the
electronic edition.}
\tablenotetext{b}{Distances from the cluster centers are given in degrees.}
\tablenotetext{c}{A "*" denotes that we have used a spectral type from the litterature in our analysis.}
\tablenotetext{d}{References correspond to: (1) Schild 1965; (2) Johnson
\& Morgan 1968; (3) Slettlebak 1968; (4) Appenzeller 1967; (5) Bidelman 1947b: (6) Morgan, Code, \& Whitford 1955. }
\tablenotetext{e}{Our cross-identifications are not complete and focus on the brighter stars 
and those for which we have spectral types.  
When available, HD or BD numbers are given.  All other IDs are taken from Wildey 1964.}
\end{deluxetable}

\clearpage
\begin{deluxetable}{ccccccc}
\tablewidth{0pc} 
\tablecolumns{7}
\tablenum{2A}
\tablecaption{Derived Quantities for Probable Cluster Members\tablenotemark{a}}
\tablehead{
  \colhead{ID} &
  \colhead{HRD\tablenotemark{b}} &
  \colhead{log T$_{eff}$} &
  \colhead{$E(B-V)$} &
  \colhead{M$_V$} &
  \colhead{M$_{bol}$} &
  \colhead{Mass [M$_\odot$]} 
}
\startdata
   3 & s &   4.300 &   0.58 &  -7.09 &  -9.09 &   33.1 \\ 
   4 & s &   4.300 &   0.63 &  -7.11 &  -9.11 &   33.1 \\ 
   9 & s &   4.385 &   0.57 &  -5.57 &  -7.99 &   21.0 \\ 
  12 & s &   4.370 &   0.42 &  -4.86 &  -7.24 &   16.4 \\ 
  16 & s &   4.340 &   0.56 &  -5.12 &  -7.34 &   16.8 \\ 
...&...&...&...&...&...&...\\ 
1582 & p &   4.144 &   0.59 &   0.59 &  -0.61 &    4.0 \\ 
\enddata
\tablenotetext{a}{The complete version of this table will be published in
the electronic version.}
\tablenotetext{b}{This column indicates how the star was placed in
the HR diagram, with an ``s" or ``p"  meaning using spectra or just
photometry, respectively. An ``a" indicates that the mean $E(B-V)$ was
adopted.}
\end{deluxetable}

\begin{deluxetable}{ccccccc}
\tablewidth{0pc} 
\tablecolumns{7}
\tablenum{2B}
\tablecaption{Derived Quantities for Probable Field Stars\tablenotemark{a}}
\tablehead{
  \colhead{ID} &
  \colhead{HRD\tablenotemark{b}} &
  \colhead{log T$_{eff}$} &
  \colhead{$E(B-V)$} &
  \colhead{M$_V$} &
  \colhead{M$_{bol}$} &
  \colhead{Mass [M$_\odot$]} 
}
\startdata
   1 & s &   3.680 &   0.16 &  -6.31 &  -6.68 &   13.7 \\ 
   2 & s &   4.000 &   0.50 &  -6.93 &  -7.20 &   16.2 \\ 
   5 & s &   4.050 &   0.73 &  -7.13 &  -7.71 &   18.8 \\ 
   6 & s &   3.940 &   0.75 &  -6.73 &  -6.77 &   13.9 \\ 
   7 & s &   3.525 &   0.62 &  -5.93 &  -7.43 &   13.6 \\ 
...&...&...&...&...&...&...\\ 
1610 & p &   4.138 &   0.62 &   0.52 &  -0.65 &    4.0 \\
\enddata
\tablenotetext{a}{The complete version of this table will be published in
the electronic edition.}
\tablenotetext{b}{This column indicates how the star was placed in 
the HR diagram, with an ``s" or ``p"  meaning using spectra or just
photometry, respectively. An ``a" indicates that the mean $E(B-V)$ was
adopted.}
\end{deluxetable}

\clearpage

\begin{deluxetable}{ccrrcrr}
\tablecolumns{5}
\tablewidth{0pc}                                  
\tablenum{3}                                    
\tablecaption{PDMF Data\label{tbl:data3a}}
\tablehead{
  \colhead{Mass Range} & 
  \colhead{} &
  \multicolumn{2}{c}{h Per} &
  \colhead{} &
  \multicolumn{2}{c}{$\chi$ Per} \\
  \cline{3-4}
  \cline{6-7} 
  \colhead{[$\cal M_\odot$]} &
  \colhead{} &
  \colhead{N} &
  \colhead{log $\xi$} &
  \colhead{} &
  \colhead{N} &
  \colhead{log $\xi$}  
}
\startdata
 4.0-5.0 && 45 & 6.80 && 37 & 6.71 \\  
 5.0-6.3 && 33 & 6.66 && 23 & 6.50 \\ 
 6.3-7.9 && 28 & 6.59 && 26 & 6.56 \\
 7.9-10.0 && 25 & 6.54 && 13 & 6.26 \\
 10.0-12.6 && 11 & 6.18 && 9 & 6.10 \\
 12.6-15.8 && 10 & 6.14 && 10 & 6.14 \\
 15.8-40.0 && 7 & 5.39 && 2 & 4.84  \\
\enddata
\end{deluxetable}

\end{document}